# Dynamic Switched Quantum Key Distribution Network with PUF-based authentication

P. Konteli *[1], N. Makris [1], E. N. Sassalou [2], S. A. Kazazis [2], A. Papageorgopoulos [1], S. Vasileiadis [3,4], K. Tsimvrakidis [1], S. Tsintzos [2], G. M. Nikolopoulos [5,6], G. T. Kanellos [1]

[1] *National and Kapodistrian University of Athens, Athens, Greece*
[2] *QUBITECH, Athens, Greece*
[3] *UBITECH, Athens, Greece*
[4] *University of Trento, Trento, Italy*
[5] *Institute of Electronic Structure and Laser, Foundation for Research and Technology-Hellas (FORTH), Heraklion, Greece*
[6] *Center for Quantum Science and Technologies, FORTH Heraklion, Crete, Greece*
*perskonteli@di.uoa.gr

**Abstract:** We demonstrate a centrally controlled dynamic switched-QKD network, with integrated PUF-based dynamic authentication for each QKD link. The performance of the dynamic switched-QKD network with real-time PUF-based authentication is analyzed. © 2025 The Author(s)

## 1. Introduction

While quantum key distribution (QKD) is considered the most secure countermeasure against quantum computers, initial authentication of QKD devices remains a major issue, especially for large-scale quantum networks [1]. Initial authentication phase is the main vulnerability for man-in-the-middle (MitM) attacks [2], in which an Eve adversary could impersonate legitimate partners, compromising the secure communication since one of the genuine peers can establish a secure key with an invalid entity [3]. To counter this problem the vast majority of QKD vendors perform initial authentication in their commercial systems using pre-shared keys (PSK). This approach is, in principle, viable for static, point-to-point (P2P) deployments but becomes increasingly impractical as QKD networks scale [1] and new nodes need to be deployed in the network. The authentication problem becomes even more complex in the case of dynamic QKD networks, where links are dynamically setup using all-optical switches between pairs of Alices and Bobs to match QKD network requirements [4] and each link requires a new authentication key as shown in Fig. 1. Other approaches include Post Quantum Cryptography (PQC) which uses mathematically hard problems that remain secure even against quantum computers [5] but is not information theoretically secure (ITS), and Physical Unclonable Functions (PUF) which exploit unique hardware characteristics to create unclonable device fingerprints [6].

To this end, a recent theoretical scheme for integrating PUF-based authentication (PBA) prior to initiating quantum key generation has been proposed in [7], arguing that PBA is preferable to other existing methods of authentication due to its ITS properties, resistance to cloning, and the elimination of the need for additional PSKs. In [8], software-defined (SDN) framework is demonstrated, specifically for managing QKD networks with PBA. However, both the QKD nodes and PUF components are software-simulated, enabling PBA and orchestration workflows without physical quantum or optical hardware. Additionally, in an experimental deployment of PUFs, authors of [9] used correlated PUFs to mitigate MitM attacks by inserting a randomized phase shift in the basis selection of the transmitted qubits during the execution of the QKD protocol, thereby securing every generated key.

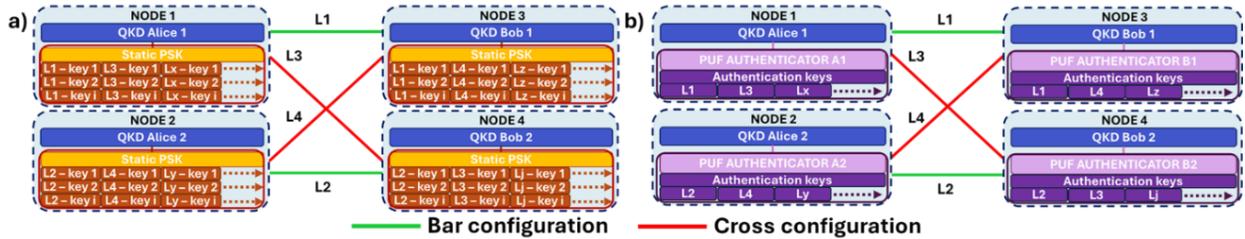

Fig. 1. Four-node dynamic switch-QKD network topology with a) PSK authentication and b) PBA schemes.

In this work, we propose for the first time the use of PBA scheme as a scalable secure solution for the initial authentication of the QKD devices in large dynamic switched-QKD networks [10], where any QKD transmitter (Alice) can arbitrarily be matched with any QKD receiver (Bob). In a PSK scheme, the authentication process for N nodes would require more than $N^2$ PSKs to authenticate each QKD pair at least one time, whereas in the PBA solution each node only needs one PUF-based authenticator for generating a new key for every new authentication link, as illustrated in Fig. 1. Scalability in the authentication of dynamically switched QKD networks is achieved by implementing a protocol for generating PBA keys, as described in [11], enabling the creation of new authentication keys for all existing active links whenever a switching operation occurs, and QKD communication must be re-established. To experimentally verify the concept, we implement a four-node switched QKD network with PBA where the switching operation between two QKD pairs is fully automated via a Python-based solution, triggering the PBA and dynamically reconfiguring all the required network parameters to transition between bar and cross configurations. We verified continuous operation by monitoring the buffer levels for each link as quantum-safe applications were running on top of the PBA-based QKD network.

## 2. Experimental setup

The experimental testbed included a 4-node switched QKD network with PBA. The authentication between each QKD link is established after every switching operation by employing a dedicated PUF authenticator for each QKD device as shown in Fig. 1(b). To implement this network, we utilized two QUKY QKD pairs by ThinkQuantum and four PUF authenticators developed by Qubitech [11]. The QUKY devices implement the three-state BB84 protocol with polarization-based encoding and a nominal power budget of 20 dB. The QKD modules use the iPOGNAC encoder with Qubit Synchronization [12], requiring only one ethernet connection for the service channel, while the quantum channels were operating at 1549.32 nm (ITU Channel 35) and 1550.92 nm (ITU Channel 33), respectively. QUKY also provide two key features that enable flexible network deployment. Firstly, they support dynamic operation through the switching process, where any Alice can connect with any Bob, while the Key Management System (KMS) remains active for all the links in the network. Secondly, they allow the manual definition of the initial authentication keys. The employed PUF-based instances generate ephemeral authentication keys by exploiting the uncontrollable manufacturing variations present in static random-access memory (SRAM) modules. The SRAM-PUF constructions considered here, utilize a RISC-based microcontroller to host the Cypress CY62256NLL-70PC SRAM module, while an ARM-based SoC serves as a PUF controller issuing challenges and validating the derived keys during the enrollment and authentication process. In the scheme illustrated in Fig. 2(a), Alice PUFs (A1 and A2) act as verifiers, while Bob PUFs (B1 and B2) act as provers, allowing the two PUFs to mutually authenticate each other.

The deployed network features fully automated and dynamic switch execution, enabled with the use of python interfaces. In particular, a python routine is executed by a central external QKD server that controls the switching operation, changing between bar and cross configuration, while this operation triggers the PBA, as shown in Fig. 2(b). The QUKY devices integrate a KMS which orchestrates all internal operations for the applied QKD configurations. At the KMS level, each device provides two independent key buffers for storing quantum-generated keys in the switched topology. Two buffers store keys produced in bar configuration for links L1 and L2, and the others store keys produced in cross configuration for links L3 and L4, as presented in Fig. 1. All buffers keep the generated keys available in all configurations, although the ones receiving new quantum keys are considered "ACTIVE", while the others "PASSIVE". That feature offers the ability to third-party encryptors to fetch keys from either buffer at any time as it is standardized in ETSI GS QKD 014 [13], and was demonstrated in this setup through a quantum-safe video conference over an IPsec tunnel secured with quantum keys, implemented using FortiGate NGFWs.

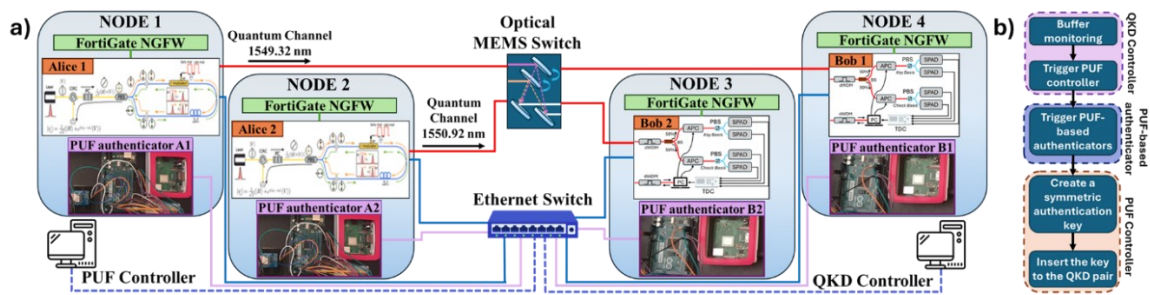

Fig. 2. a) Topology of the 4-node QKD network with the PUF-based authenticators and b) an abstracted flowchart of the monitoring and the switching operation of the network controllers.

## 3. Results and Discussion

To validate the experimental setup, we let the architecture operate for approximately 50 minutes during which switching operations occurred dynamically based on the network conditions, specifically the buffers level. To emulate buffer depletion, a bash script was executed to retrieve keys from the buffers though ETSI 014 requests [13], with a rate of 1 key/sec for L1 and L2 and a rate of 0.2 keys/sec for L3 and L4. Along with the bash script, Layer 3 consumers were fetching keys from the passive buffers with a rate of 0.5 keys/minute exhibiting a mean latency of 114.7 msec and jitter of 18.6 msec. The QKD controller prioritized the buffers of bar configuration, so the switching threshold was set to 80,000 kB, whereas for the cross (non-prioritized) link it was set to 60,000 kB. Whenever the respective threshold conditions were met, the network switched between configurations to refill the corresponding buffers as shown in Fig. 3(c), (g). Fig. 3(a) illustrates the status of the buffers indicating which buffer was active each time.

During the switching operation all buffers are deactivated, setting their status to "PASSIVE", and the generation of quantum keys is stopped. The switching procedure is considered complete when the buffers are reactivated and the Base-agreement phase is initiated for both QKD pairs. Each time a switching operation occurred, a new initial PBA key was generated for every active link. The detailed timing results for both the PBA and the switching operation are presented in Table 1. The average total verification time for both PUFs was approximately 27 seconds with the PUF-based key generation lasting ~7 seconds, while the total switching process had an average duration of 123.5 seconds, as measured after four switching operations. Fig. 3(b), (d), (e), (f), and (h) illustrate the Secure Key Rate (SKR), Raw Key Rate (RKR), and Quantum Bit Error Rate (QBER) for all links. The mean SKR produced for each link is 2816.2 bps for L1, 2727.3 bps for L2, 3914.9 bps for L3 and 2304.4 bps for L4.

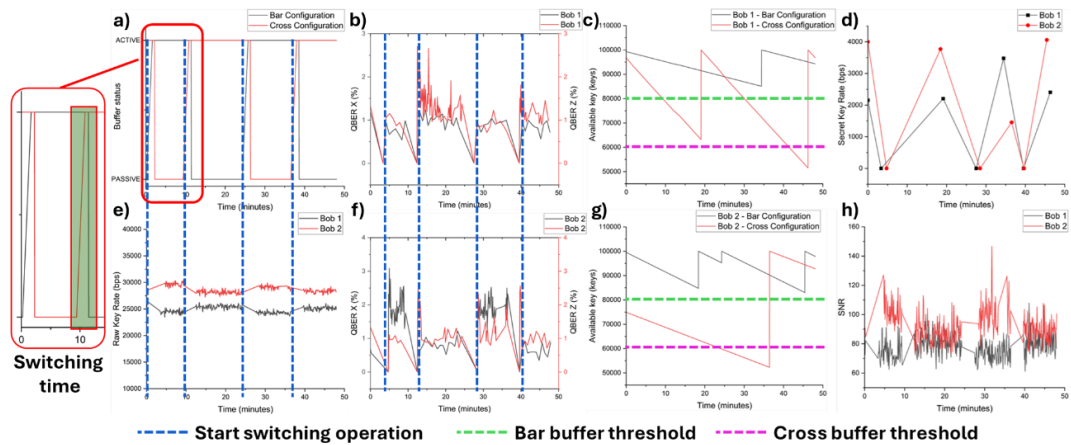

Fig. 3. Key performance indicators for network operation. a) Buffer status, b), f) QBER and c) buffer levels for L1 and L4 and g) buffer levels for L2 and L3 and d), e), h) SKR, RKR and SNR for all links.

Table 1. Average time for each component action of the PUF-based authenticator and Total switching time

| | Time Measurements (s) | | | | | | | | |
|---|---|---|---|---|---|---|---|---|---|
| | A1 | A2 | B1 | B2 | | A1 | A2 | B1 | B2 |
| PUF HTTPS request | 12.9 | 12.9 | - | - | SSH processes | 7.1 | 7.1 | 6.1 | 6.1 |
| Arduino interaction | 7 | 6.7 | 6.8 | 6.7 | Verification process | 27 | 26.7 | - | - |
| Hashing processes | 0.002 | 0.002 | - | - | Switching Operation | 123.5 | | | |

## 4. Conclusion

This work demonstrated a 4-node switched QKD network integrating PBA, enabling secure and dynamic reconfiguration of quantum links. The experimental results validated the successful operation of automated switching and authentication processes across all configurations. The integration of SRAM-based PUFs provided a hardware-rooted mechanism for establishing initial authentication keys without relying on PSKs. Overall, the proposed architecture showcases a scalable and quantum-secure approach for large dynamic QKD network deployments.

## 5. Acknowledgements

This work was funded by the EC HellasQCI (GA 101091504) and by the EU project QSNP (GA 101114043).